\documentclass{aastex62}

\usepackage{hyperref}
\usepackage{multirow}
\usepackage{amsmath}
\usepackage{graphicx}

\shorttitle{Predicting CMEs Using \textit{SDO}/HMI Vector Magnetic Data Products and RNNs}
\shortauthors{Liu et al.}

\begin{document}

\title{{\bf \large Predicting Coronal Mass Ejections Using \textit{SDO}/HMI Vector Magnetic Data Products and Recurrent Neural Networks}}

\author{Hao Liu}
\affiliation{Institute for Space Weather Sciences, New Jersey Institute of Technology, University Heights, Newark, NJ 07102-1982, USA 
hl422@njit.edu, chang.liu@njit.edu, wangj@njit.edu, haimin.wang@njit.edu}
\affiliation{Department of Computer Science, New Jersey Institute of Technology, University Heights, Newark, NJ 07102-1982, USA}

\author{Chang Liu}
\affiliation{Institute for Space Weather Sciences, New Jersey Institute of Technology, University Heights, Newark, NJ 07102-1982, USA 
hl422@njit.edu, chang.liu@njit.edu, wangj@njit.edu, haimin.wang@njit.edu}
\affiliation{Big Bear Solar Observatory, New Jersey Institute of Technology, 40386 North Shore Lane, Big Bear City, CA 92314-9672, USA}
\affiliation{Center for Solar-Terrestrial Research, New Jersey Institute of Technology, University Heights, Newark, NJ 07102-1982, USA}

\author{Jason T. L. Wang}
\affiliation{Institute for Space Weather Sciences, New Jersey Institute of Technology, University Heights, Newark, NJ 07102-1982, USA 
hl422@njit.edu, chang.liu@njit.edu, wangj@njit.edu, haimin.wang@njit.edu}
\affiliation{Department of Computer Science, New Jersey Institute of Technology, University Heights, Newark, NJ 07102-1982, USA}

\author{Haimin Wang}
\affiliation{Institute for Space Weather Sciences, New Jersey Institute of Technology, University Heights, Newark, NJ 07102-1982, USA 
hl422@njit.edu, chang.liu@njit.edu, wangj@njit.edu, haimin.wang@njit.edu}
\affiliation{Big Bear Solar Observatory, New Jersey Institute of Technology, 40386 North Shore Lane, Big Bear City, CA 92314-9672, USA}
\affiliation{Center for Solar-Terrestrial Research, New Jersey Institute of Technology, University Heights, Newark, NJ 07102-1982, USA}

\begin{abstract}
We present two recurrent neural networks (RNNs),
one based on gated recurrent units and the other based on long short-term memory,
for predicting whether an active region (AR) that produces an M- or X-class flare  
will also produce a coronal mass ejection (CME).
We model data samples in an AR as time series and use the RNNs to capture temporal information of the data samples. 
Each data sample has 18 physical parameters, or features, 
derived from photospheric vector magnetic field data taken by
the Helioseismic and Magnetic Imager (HMI)
on board the \textit{Solar Dynamics Observatory} (\textit{SDO}). 
We survey M- and X-class flares that occurred from 2010 May to 2019 May 
using the \textit{Geostationary Operational Environmental Satellite}'s X-ray flare catalogs 
provided by the National Centers for Environmental Information (NCEI), 
and select those flares with identified ARs in the NCEI catalogs. 
In addition, we extract the associations of flares and CMEs
 from the Space Weather Database Of Notifications, Knowledge, Information (DONKI).
We use the information gathered above to build the labels (positive versus negative) of the data samples at hand.
Experimental results demonstrate the superiority of our RNNs over closely related machine learning methods
in predicting the labels of the data samples.
We also discuss an extension of our approach to predict a probabilistic estimate of 
how likely an M- or X-class flare will initiate a CME, with good performance results.
To our knowledge this is the first time that RNNs have been used for CME prediction.
\end{abstract}

\keywords{Sun: activity $-$ Sun: flares $-$ Sun: coronal mass ejections (CMEs)}

\section{Introduction} \label{sec:intro}

Coronal mass ejections (CMEs) are intense bursts of magnetic flux and plasma 
that are ejected from the Sun into interplanetary space \citep{2000JGR...105.2375L}. 
They are often associated with solar flares and originated from active regions (ARs) on the Sun's photosphere 
where magnetic fields are strong and evolve rapidly. 
Major CMEs and their associated flares can cause severe influences on the near-Earth environment, 
resulting in potentially life-threatening consequences \citep{Baker2004}. 
Therefore, substantial efforts are being invested on developing new technologies 
for early detection and forecasting of flares and CMEs \citep{Bobra2016CME,Inceoglu2018}.

Both flares and CMEs are believed to be magnetically-driven events; 
evidence shows that they may constitute different manifestations of the same physical process 
\citep{1995A&A...304..585H, 2012ApJ...753...88B, Gosling2013}. 
However, solar observations over the past few decades have 
clearly indicated that there may not be a one-to-one
correspondence between flares and CMEs, and their relationship is still under active investigation 
\citep[see, e.g.,][]{2009IAUS..257..233Y, 2018ApJ...869...99K}. 
Much effort has been devoted to analyzing the structural properties of coronal magnetic fields, 
which may play an important role in determining whether an eruption evolves into a CME or remains as a confined flare 
\citep{Torok2005, DeVore&Antiochos2008, LiuY2008, Baumgartner2018}. 
In the meantime, many researchers have investigated the relationship 
between CME productivity and the features of the photospheric magnetic field of flare-productive ARs 
where the features can be directly derived from photospheric vector magnetograms. 
For example, \citet{Qahwaji2008} used properties such as flare duration along with machine learning algorithms
to predict whether a flare is likely to initiate a CME.
\citet{Bobra2016CME} used 18 physical features, or predictive parameters, derived from the photospheric vector magnetic field data 
provided by the Helioseismic and Magnetic Imager \citep[HMI;][]{2012SoPh..275..229S} 
on board the \textit{Solar Dynamics Observatory} \citep[\textit {SDO};][]{2012SoPh..275....3P} 
to forecast whether a CME would be associated with an M- or X-class flare.
The flares are classified according to the peak flux
 (in watts per square meter, $W/m^{2}$) of 1 to 8 \AA ~X-rays as
measured by the \textit {Geostationary Operational Environmental Satellite} (\textit{GOES}).
These authors found that using a combination of six intensive parameters 
captured most of the relevant information contained in the photospheric magnetic field. 
\citet{Inceoglu2018} later developed methods to forecast whether flares 
would be associated with CMEs and solar energetic particles (SEPs).
The authors employed two machine learning algorithms, 
support vector machines (SVMs) and multilayer perceptrons (MLPs),
and showed that SVMs performed slightly better than MLPs in the forecasting task.

Machine learning is a non-physics-based technology used in predictive analytics.
It is a subfield of artificial intelligence, which grants computers abilities to learn from the past data and 
make predictions on unseen future data \citep{Alpaydin-2016,DBLP:books/daglib/0040158}. 
Many different machine learning-based techniques have been developed for solar flare prediction
\citep[see, e.g.,][]{2017ApJ...843..104L, 2018SoPh..293...28F, 2018SoPh..293...48J, 2018ApJ...858..113N, Liu2019}. 
However, CME prediction has been mainly based on
SVMs \citep{Bobra2016CME}
and MLPs \citep{Inceoglu2018}.

In this paper, we attempt to extend the work of \citet{Bobra2016CME}
by proposing new machine learning algorithms and applying the algorithms to
 {\itshape SDO}/HMI vector magnetic field data 
 to predict  whether an AR that produces 
an M- or X-class flare will also produce a CME.
The machine learning algorithms we explore here include two kinds of recurrent neural networks \citep[RNNs;][]{Hopfield2554}:
a long short-term memory (LSTM) network \citep{Hochreiter1997} and a gated recurrent unit (GRU) network \citep{Cho2014}. 
LSTM cells and GRUs differ in the number and type of gates employed in the networks---an LSTM cell has three gates (input, output and forget gates) whereas
a GRU has two gates (reset and update gates).
RNNs can use their internal state (memory) and gates to process sequences of inputs, 
which makes them suitable for tasks 
such as speech recognition, handwriting recognition and time series forecasting \citep{LeCun2015, DBLP:books/daglib/0040158}. 
In a CME prediction task, the observations in each AR form time series data, and
hence RNNs work well in this task. 
To our knowledge, this is the first time that RNNs are used for CME prediction. 

The rest of this paper is organized as follows. 
Section $2$ describes our data collection scheme
and predictive parameters used in the study.
Section 3 presents our RNN architectures and algorithms.
Section 4 reports experimental results.
Section 5 concludes the paper.  

\section{Data and Predictive Parameters}
\label{sec:data}
We adopt the data products, named Space-weather HMI Active Region Patches \citep[SHARP;][]{2014SoPh..289.3549B}, 
produced by the \textit{SDO}/HMI team. 
These data were released at the end of 2012 
\citep{2014SoPh..289.3549B} 
and can be found in the \textsf{hmi.sharp} data series at 
the Joint Science Operations Center 
(JSOC).\footnote{\url{http://jsoc.stanford.edu/}}
The SHARP data series contains ARs tracked throughout their lifetime 
and provides many physical parameters suitable for flare/CME predictions. 

We collect data samples at a cadence of 12 minutes where the data samples are retrieved from 
the \textsf{hmi.sharp\_cea\_720s} 
definitive data series on the JSOC website by using SunPy \citep{2015CS&D....8a4009S}. 
We use the same 18  features, or SHARP parameters, as described in Table 1  of \citet{Bobra2016CME} 
that characterize AR magnetic field properties for CME prediction. 
These 18 features or predictive parameters include
MEANPOT (mean photospheric magnetic free energy), 
SHRGT45 (fraction of area with shear $> 45^\circ$), 
TOTPOT (total photospheric magnetic free energy density), 
USFLUX (total unsigned flux), 
MEANJZH (mean current helicity), 
ABSNJZH (absolute value of the net current helicity), 
SAVNCPP (sum of the modulus of the net current per polarity), 
MEANALP (mean characteristic twist parameter),
MEANSHR (mean shear angle), 
TOTUSJZ (total unsigned vertical current), 
TOTUSJH (total unsigned current helicity), 
MEANGAM (mean angle of field from radial), 
MEANGBZ (mean gradient of vertical field), 
MEANJZD (mean vertical current density), 
AREA\_ACR (area of strong field pixels in the active region), 
R\_VALUE (sum of flux near polarity inversion line), 
MEANGBT (mean gradient of total field) and 
MEANGBH (mean gradient of horizontal field).
Because the features have different units and scales, we normalize the feature values as done in \citet{Liu2019}.
Data samples with incomplete features are excluded from our dataset \citep{Bobra2016CME}.

Our proposed RNNs require labeled training samples. 
We survey M- and X-class flares that occurred in the period between 2010 May and 2019 May, 
using the \textit{GOES} X-ray flare catalogs provided by the National Centers for Environmental Information (NCEI), 
and select M- and X-class flares with identified ARs in the NCEI flare catalogs. 
As in \citet{Bobra2016CME},
flares that are outside $\pm$ $70^\circ$ of the central meridian during the \textit{GOES} X-ray flux peak time
are excluded from our dataset.
Flares where the (1) absolute value of the radial velocity of \textit{SDO} is larger than 3500 m~s$^{-1}$ or
(2) HMI data are of low quality as described in \citet{Hoeksema2014}
are also excluded from our dataset. 
In this way, we exclude data with noise or low quality,
and keep data of high quality in our study.
In addition, we extract data from a NASA Space Weather Research Center database 
called Space Weather Database Of Notifications, Knowledge, Information (DONKI)\footnote{\url{http://kauai.ccmc.gsfc.nasa.gov/DONKI/}} 
to label whether or not any given flare produced a CME. 
This yields a database of 129 M- and X-class flares that are associated with CMEs and 
610 M- and X-class flares that are not associated with CMEs. 

\section{Methodology}\label{sec:methodology}
\subsection{Prediction Task}
\label{Task}
Following \citet{Bobra2016CME}, 
we intend to use past observations of a flaring AR to predict its future CME productivity. 
Specifically, we want to solve the following binary classification problem: 
will an AR that produces an M- or X-class flare within the next $T$ hours 
also produce a CME associated with the flare? 
We consider $T$ ranging from 12 to 60 
 in 12 hr intervals. 
These prediction times are commonly used by researchers
\citep{Ahmed2013, Bobra2016CME, Inceoglu2018}.

Our dataset contains data samples collected within the $T$ hours prior to the peak time of an M- or X-class flare regardless of
whether the flare produces a CME.
Those data samples collected within the $T$ hours prior to the peak time of an M- or X-class flare 
that produces a CME belong to the positive class;
the other data samples belong to the negative class.
Data samples collected in years 2010--2014 are used for training, and those in years 2015--2019 are used for testing. 
Thus, the training set and test set are disjoint, and hence our proposed RNNs will make predictions on ARs that they have never seen before. 
Table \ref{tab:numSamples} summarizes the numbers of positive and negative samples for different $T$ values 
used for training and testing respectively.

If a data sample is missing at some time point or 
if there are insufficient data samples within the $T$ hours prior to the peak time of an M- or X-class flare,
we adopt a zero-padding strategy by adding synthetic data samples with zeros for all feature values 
to yield a complete, non-gapped time-series dataset. 
This zero-padding method is used after normalizing the feature values.
Therefore, the zero-padding method does not affect the normalization procedure.
For a given time point $t$ and an AR that produces an M- or X-class flare within the next $T$ hours of $t$, 
the proposed RNNs predict whether or not the flare will initiate a CME.

\begin{table}[]
	\centering
	\caption{Numbers of Positive and Negative Samples Collected for Different Hours Used in This Study}
	\label{tab:numSamples}
	\begin{tabular}{c||cc||cc||cc||cc||cc}
		\hline
		\hline
		\multirow{2}{*}{} & \multicolumn{2}{c||}{12 hr} & \multicolumn{2}{c||}{24 hr} & \multicolumn{2}{c||}{36 hr} & \multicolumn{2}{c||}{48 hr} & \multicolumn{2}{c}{60 hr} \\ \cline{2-11} 
		& Positive   & Negative  & Positive   & Negative  & Positive   & Negative  & Positive   & Negative & Positive   & Negative \\ \hline
		Training          &  3,387 &  16,960    &  6,323  &  27,281 &  9,109 &  34,994  & 11,613 & 41,402  & 13,775  &  46,994  \\ 
		Testing           &  550   &  762    &   922  & 1,059 & 1,214  &  1,253  & 1,540 & 1,360  & 1,814 & 1,483  \\
		\hline
	\end{tabular}
\end{table}

\subsection{Prediction Methods}
\begin{figure}
	\centering
	\includegraphics[width=3.5in, trim=4 2 4 2,clip]{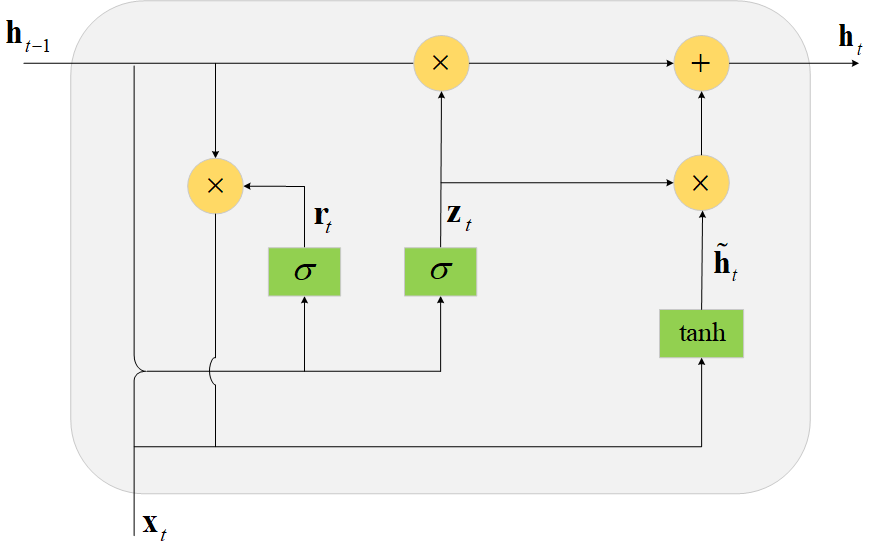}   \\
	\caption{Illustration of a GRU where 
		$\textbf{z}_t$ is the update gate, 
		$\textbf{r}_t$ is the reset gate,
		$\textbf{h}_t$ is the output vector (memory content) and
		$\textbf{x}_{t}$ is the input vector to the GRU.
                   The subscript $t$ indexes the time step.
}
	\label{fig:gru} 
\end{figure}

We employ two kinds of RNNs: a GRU network \citep{Cho2014, DBLP:books/daglib/0040158} 
and an LSTM network \citep{Hochreiter1997, DBLP:books/daglib/0040158}.
A GRU contains three interactive parts including a memory content, an update gate and a reset gate, as illustrated in Figure \ref{fig:gru}
where boldface is used for matrix notations.
The new memory content $\textbf{h}_t$ is updated by the previous memory content $\textbf{h}_{t-1}$ 
and the candidate memory content $\tilde{\textbf{h}}_t$ as follows \citep{Cho2014, DBLP:books/daglib/0040158}: 
\begin{equation}
\textbf{h}_t = \textbf{z}_t \odot \textbf{h}_{t-1} + (1-\textbf{z}_t) \odot \tilde{\textbf{h}}_t,
\end{equation}
where the update gate $\textbf{z}_t$ that determines how much of the past information from previous time steps 
needs to be passed to the future is calculated as follows \citep{Cho2014, DBLP:books/daglib/0040158}:
\begin{equation}
\textbf{z}_t = \sigma (\textbf{W}_z \cdot [\textbf{h}_{t-1}, \textbf{x}_t] + \textbf{B}_z),
\end{equation} 
and the reset gate $\textbf{r}_t$ that determines how much of the past information to forget 
is computed as follows \citep{Cho2014, DBLP:books/daglib/0040158}:
\begin{equation}
\textbf{r}_t = \sigma (\textbf{W}_r \cdot [\textbf{h}_{t-1}, \textbf{x}_t] + \textbf{B}_r).
\end{equation}
Here $\textbf{x}_t$ represents the input vector at time step $t$. 
The candidate memory content $\tilde{\textbf{h}}_t$ is computed as follows 
\citep{Cho2014, DBLP:books/daglib/0040158}:
\begin{equation}
\tilde{\textbf{h}}_t = \mbox{tanh}(\textbf{W}_h \cdot [\textbf{r}_t \odot \textbf{h}_{t-1}, \textbf{x}_t] + \textbf{B}_h).
\end{equation}
In the above equations, $\textbf{W}$ and $\textbf{B}$ contain weights and biases respectively, which need to be learned during training; 
$[.]$ denotes the concatenation of two vectors;
$\sigma(\cdot)$ is the sigmoid function, 
i.e., $\sigma(\textbf{z})=\frac{1}{1+e^{-\textbf{z}}}$; 
$\mbox{tanh}(\cdot)$ is the hyperbolic tangent function,
i.e., $\mbox{tanh}(\textbf{z})=\frac{e^\textbf{z}-e^{-\textbf{z}}}{e^\textbf{z}+e^{-\textbf{z}}}$; 
$\odot$ denotes the Hadamard product (element-wise multiplication).

Our GRU network contains a GRU layer with $m$ GRUs (in the study presented here, $m$ is set to 20). 
We add an attention layer with $m$ neurons above the GRU layer
to focus on information in relevant time steps as done in \citet{Liu2019}.
We then add a fully connected layer with 100 neurons on top of the attention layer.
Finally, the output layer with one neuron, which is activated by the sigmoid function, produces predicted values. 
Our LSTM network is similar to the GRU network except that
the GRU layer is replaced by an LSTM layer with $m$ LSTM cells.
This is reminiscent of the LSTM network presented in \citet{Liu2019}.

Let $\textbf{x}_{t}$ represent the data sample collected at time point $t$.
During training, for each time point $t$,
we take $m$ consecutive data samples
$\textbf{x}_{t-m+1}, \textbf{x}_{t-m+2}, \dots, \textbf{x}_{t-1}, \textbf{x}_t$ from the training set
and use the $m$ consecutive data samples to train the proposed RNNs
including the GRU and LSTM networks.
The label of these $m$ consecutive data samples is defined
to be the label of the last data sample $\textbf{x}_{t}$.
Thus, if $\textbf{x}_{t}$ belongs to the positive class, then the input sequence
$\textbf{x}_{t-m+1}, \textbf{x}_{t-m+2}, \dots, \textbf{x}_{t-1}, \textbf{x}_t$ 
is defined as positive; otherwise the sequence is defined as negative.
Because the data samples are collected continuously at a cadence of 12 minutes
and missing values are filled up by our zero-padding strategy,
the input sequence spans $\frac{m}{5}$ hours. 

Also during training, we use a weighted cross entropy cost function for optimizing model parameters where
the cost function is computed as follows \citep{DBLP:books/daglib/0040158}:
 \begin{equation}
\label{eq:loss}
J=\sum_{n=1}^N\omega_{0}y_{n}\textrm{log}(\hat{y}_{n})+\omega_{1}(1-y_{n})\textrm{log}(1-\hat{y}_{n}).
\end{equation}
Here, $N$ is the total number of sequences each having $m$ consecutive data samples in the training set, 
$\omega_{0}$ and $\omega_{1}$ are the weights of the positive and negative classes respectively, 
which are derived by the ratio of the sizes of the positive and negative classes with more weight given to 
the minority class.\footnote{Refer to the training data in Table \ref{tab:numSamples}. 
The minority class is the positive class.}
We use $y_{n}$ and $\hat{y}_{n}$ to denote the observed probability 
(which is equal to $1$ if the $n$th sequence belongs to the positive class) 
and the estimated probability of the $n$th sequence, respectively.

The proposed RNN methods are implemented in Python, TensorFlow and Keras. 
A mini-batch strategy \citep{DBLP:books/daglib/0040158} is used 
to achieve faster convergence during backpropagation. 
The optimizer used is RMSprop, 
which is a method for gradient descent, 
where the learning rate is set to 0.001.
The batch size is set to 256
and the number of epochs is set to 20.
The length of each input sequence, $m$, is set to 20, 
meaning that every time 20 consecutive data samples are used as input to our RNNs. 
The hyperparameter values, the optimizer, 
and the cost function in Equation (\ref{eq:loss})
are chosen to maximize the TSS scores to be defined in the experiments section.

During testing, to predict whether a given AR that produces an M- or X-class flare 
within the next $T$ hours of a time point $t$ will also produce a CME associated with the flare, 
we take $\textbf{x}_t$ and its preceding $m-1$ data samples, and then feed the $m$ consecutive test data samples 
$\textbf{x}_{t-m+1}, \textbf{x}_{t-m+2}, \dots, \textbf{x}_{t-1}, \textbf{x}_t$ into the trained RNNs. 
The output of the RNNs, i.e., the predicted result, is a scalar number with a value of 1 or 0, 
indicating $\textbf{x}_t$ is positive (i.e., the AR will also produce a CME associated with the flare) 
or $\textbf{x}_t$ is negative (i.e., the AR will not produce a CME associated with the flare). 
This value is determined by comparing the probability calculated by the sigmoid function 
in the output layer of the RNNs
with a threshold. 
If the probability is greater than or equal to the threshold, then $\textbf{x}_t$ is predicted to be positive; 
otherwise $\textbf{x}_t$ is predicted to be negative. 
It should be pointed out that, the way we use 
the $m$ consecutive test data samples
$\textbf{x}_{t-m+1}, \textbf{x}_{t-m+2}, \dots, \textbf{x}_{t-1}, \textbf{x}_t$ to
predict whether a given AR that produces an M- or X-class flare 
within the next $T$ hours of a time point $t$ will also produce a CME associated with the flare
 is different from the previously published machine learning methods
\citep{Bobra2016CME}, 
which used only the test data sample $\textbf{x}_{t}$ to make the prediction. 

\section{Results}\label{sec:results}

\subsection{Metrics and Experimental Setup}
\label{metrics}

Given an AR that produces an M- or X-class flare within the next $T$ hours of a time point $t$ and a data sample $\textbf{x}_{t}$ observed at time point $t$,
we define $\textbf{x}_{t}$ to be a true positive (TP) if 
our RNNs predict that $\textbf{x}_{t}$ is positive, 
and $\textbf{x}_{t}$ is indeed positive, i.e., the M- or X-class flare produces, or is associated with, a CME.
We define $\textbf{x}_{t}$ to be a false positive (FP) if 
our RNNs predict that $\textbf{x}_{t}$ is positive 
while $\textbf{x}_{t}$ is actually negative, i.e., the M- or X-class flare does not produce, or is not associated with, a CME. 
We say $\textbf{x}_{t}$ is a true negative (TN) if 
our RNNs predict $\textbf{x}_{t}$ to be negative
and $\textbf{x}_{t}$ is indeed negative;
$\textbf{x}_{t}$ is a false negative (FN) if 
our RNNs predict $\textbf{x}_{t}$ to be negative
while $\textbf{x}_{t}$ is actually positive.
We also use TP (FP, TN, FN, respectively) to represent the total number of 
true positives (false positives, true negatives, false negatives, respectively)
produced by our RNNs.

The performance metrics used in this study include the following:
\begin{equation}
\text{Recall} = \frac{\mbox{TP}}{\mbox{TP + FN}}, 
\end{equation}
\begin{equation}
\text{Precision} = \frac{\mbox{TP}}{\mbox{TP + FP}} , 
\end{equation}
\begin{equation}
\text{Accuracy (ACC)} = \frac{\mbox{TP + TN}}{\mbox{TP + FP + TN + FN}}, 
\end{equation}
\begin{equation}
\text{Heidke Skill Score (HSS)} = \frac{2(\mbox{TP}\times \mbox{TN}-\mbox{FP}\times \mbox{FN})}{(\mbox{TP+FN})(\mbox{FN+TN})+(\mbox{TP+FP})(\mbox{FP+TN})},
\end{equation}
\begin{equation}
\text{True Skill Statistics (TSS)} = \frac{\mbox{TP}}{\mbox{TP + FN}} - \frac{\mbox{FP}}{\mbox{TN + FP}}.
\end{equation}

The Heidke Skill Score (HSS)
\citep{heidke1926berechnung} is used to measure the fractional improvement of our prediction over the random prediction
\citep{2018SoPh..293...28F}. 
The TSS score is the recall subtracted by the false alarm rate \citep{2012ApJ...747L..41B}. 
We also use the area under the curve (AUC) in a receiver operating characteristic (ROC) curve \citep{marzban2004roc}, 
which represents the degree of separability, 
indicating how well a method is capable of distinguishing between two classes 
with the ideal value of one.
These performance metrics are commonly used when dealing with binary
classification problems such as the one at hand as defined in the beginning of Section \ref{Task}.
In general, the larger HSS, TSS and AUC score a binary classification method has,
the better performance the method achieves.

To gain a better understanding of the behavior of the proposed RNNs,
we adopt the following cross-validation methodology.
We partition the training (test, respectively) set into 10 equal-sized folds. 
For every two training (test, respectively) folds $i$ and $j$, $i \not= j$, fold $ i$ and fold $ j$ are disjoint; 
furthermore, fold $i$ and fold $j$ contain approximately the same number of positive training (test, respectively) data samples
and approximately the same number of negative training (test, respectively) data samples. 
In the $i$th run, $1 \leq i \leq 10$,
all training data samples except those in training fold $i$ are used to train a model,
and the trained model is used to make predictions on all test data samples except those in test fold $i$.
We calculate the performance metric values based on the predictions made in the $i$th run.
There are 10 runs.
The means and standard deviations over the 10 runs are calculated and recorded.

\subsection{Feature Assessment}

We conduct a series of experiments to analyze the importance of each of the 18 features studied here 
using the cross-validation methodology described above and
the feature assessment method introduced in Section 4.3 of \citet{Liu2019}. 
Each time only one feature is used to make predictions.
The probability threshold used by our RNNs is set to maximize the TSS score in each run. 
There are 10 runs and the corresponding mean TSS score is calculated and recorded. 
There are 18 features, so 18 mean individual TSS scores are recorded. 
These 18 mean individual TSS scores are sorted in descending order, 
and the 18 corresponding features are ranked from the most important (with the highest mean individual TSS score) 
to the least important (with the lowest mean individual TSS score) accordingly. 
Table \ref{tab:features} presents the 18 features ranked by our GRU and LSTM networks respectively
for different $T$ values where $T$ ranges from 12 to 60 in 12 hr intervals.
It can be seen from the table that
MEANPOT, SHRGT45, ABSNJZH and SAVNCPP are consistently ranked in the top 10 list 
by both LSTM and GRU networks.
In particular, MEANPOTS plays the most important role in CME prediction, which is ranked as the top one in all cases. 
Other features such as TOTPOT, USFLUX, MEANJZH, MEANALP and MEANSHR also show strong predictive power and
are ranked in the top 10 list in most cases.
Compared to the top 10 lists of \citet{Bobra2016CME}, who used a different feature ranking method for $T$ = 24, 48
\citep[see Table 1 of][]{Bobra2016CME},
these lists overlap to some extent 
(seven features simultaneously occur in the top 10 lists given by \citet{Bobra2016CME} and our methods).
 
\begin{table}
	\centering
	\caption{Rankings of the 18 \textit{SDO}/HMI Magnetic Parameters Used in Our Study}
	\label{tab:features}
	\begin{tabular}{lcccccccccc}
		\hline
		\hline
		SHARP & \multicolumn{2}{c}{12 hr} & \multicolumn{2}{c}{24 hr} & \multicolumn{2}{c}{36 hr} & \multicolumn{2}{c}{48 hr} & \multicolumn{2}{c}{60 hr} \\ 
		Keyword & GRU & LSTM & GRU & LSTM & GRU & LSTM & GRU & LSTM & GRU & LSTM \\ \hline
		MEANPOT & 1 & 1 & 1 & 1 & 1 & 1 & 1 & 1 &1 & 1 \\ \hline
		SHRGT45 & 2 & 2 & 2 & 2 & 3 & 3 & 3 & 4 & 3 & 4 \\ \hline
		TOTPOT & 12 & 10 & 4 & 9 & 2 & 4 & 2 & 2 & 2 & 2 \\ \hline
		USFLUX & 9 & 12 & 6 & 4 & 5 & 9 & 5 & 5 &12 & 5 \\ \hline
		MEANJZH & 5 & 5 & 3 & 5 & 7 & 7 & 12 & 8 & 6 & 7 \\
		\hline
		ABSNJZH & 3 & 4 & 8 & 8 & 9 & 8 & 8 & 7 & 5 & 6 \\ \hline
		SAVNCPP & 4 & 6 & 9 & 7 & 8 & 6 & 7 & 9 & 7 & 8  \\ \hline
		MEANALP & 7 & 7 & 7 & 10 & 6 & 10 & 9 & 12 & 10 & 12 \\ \hline
		MEANSHR &  17 & 3 & 5 & 3 & 4 & 2 & 4 & 3 & 4 & 3 \\ \hline
		TOTUSJZ & 8 & 8 & 10 & 11 & 11 & 11 & 10 & 10 & 9 & 9 \\ \hline
		TOTUSJH & 14 & 18 & 15 & 17 & 10 & 12 & 11 & 11 & 11 & 11 \\ \hline
		MEANGAM & 18 & 17 & 14 & 6 & 12 & 5 & 16 & 6  & 8 & 10 \\ \hline
		MEANGBZ & 11 & 13 & 11 & 12 & 14 & 14 & 14 & 14 & 18 & 14 \\ \hline
		MEANJZD & 6 & 9 & 12 & 13 & 16 & 15 & 17 & 15 & 16 & 15 \\ \hline
		AREA\_ACR & 10 & 11 & 13 & 14 & 15 & 16 & 16 & 17 & 14 & 16 \\ \hline
		R\_VALUE & 15 & 16 & 16 & 15 & 13 & 13 & 13 & 13 & 13 & 13 \\ \hline
		MEANGBT & 13 & 14 & 17 & 16 & 17 & 17 & 15 & 16 & 17 & 17 \\ \hline
		MEANGBH & 16 & 15 & 18 & 18 & 18 & 18 & 18 &  18 & 15 & 18 \\ \hline
	\end{tabular}
\end{table} 

Next, according to the ranked features, mean cumulative TSS scores are calculated. 
Specially, the mean cumulative TSS score of the top $k$, $1 \leq k \leq 18$, most important features 
is equal to the mean TSS score of using the top $k$ most important features altogether for CME prediction. 
We calculate the mean cumulative TSS scores for $T=$12, 24, 36, 48 and 60 hours. 
It is found that using all the 18 features together does not yield the highest mean cumulative TSS scores. 
In fact, using the top 16, 12, 9, 14 and 5 features for $T=$ 12, 24, 36, 48 and 60 hours respectively yields the highest mean cumulative TSS scores, 
achieving the best performance for our GRU network. 
Using the top 15, 12, 8, 15 and 6 features for $T=$12, 24, 36, 48, 60 hours respectively yields the highest mean cumulative TSS scores for our LSTM network. 
This happens probably because low ranked features are noisy features, 
and using them may deteriorate the performance of the methods.
In subsequent experiments, we use the best features for each method. 

\subsection{Performance Comparison}

We compare our proposed RNNs with three closely related machine learning methods including 
a multilayer perceptron (MLP) \citep{Inceoglu2018},
a support vector machine (SVM) \citep{Bobra2016CME} and
the random forest algorithm (RF) \citep{2017ApJ...843..104L}. 
MLP and SVM have been previously used for CME prediction \citep{Bobra2016CME,Inceoglu2018}.
RF has been used in flare prediction with good performance
\citep{2017ApJ...843..104L, 2018SoPh..293...28F,  Liu2019}.
These three machine learning methods are inherently probabilistic forecasting models \citep{2018SoPh..293...28F} in the sense that
each of them predicts a probability.
Since we are dealing with a binary classification problem, as defined in the beginning of Section \ref{Task},
we convert each probabilistic forecasting model into a binary classification model \citep{2018ApJ...858..113N, 2018SoPh..293...48J}
by comparing the predicted probability with a threshold.
If the predicted probability is greater than or equal to the threshold, 
then the model predicts that a flare will produce, or is associated with, a CME; 
otherwise, the model predicts that the flare will not produce a CME.

MLP consists of an input layer, an output layer and two hidden layers both with 200 neurons. 
SVM uses the radial basis function (RBF) kernel. 
RF has two parameters: mtry (number of features randomly selected to split a node) 
and ntree (number of trees to grow in the forest). 
We vary the values of ntree $\in$ \{300, 500, 1,000\} and mtry $\in$ [2, 8], 
and set ntree to 500 and mtry to 3. 
The hyperparameter and parameter values 
used by these three related machine learning methods
are chosen to maximize their TSS scores.
As in the proposed RNNs, we use 
data samples collected in years 2010-2014 for training 
and data samples in years 2015-2019 for testing.
To deal with the imbalanced datasets described in Table \ref{tab:numSamples}, 
we give more weight to the minority class during training 
as done for the RNNs. 
The same cross-validation methodology as described in Section \ref{metrics}
is adopted to evaluate the performance of the three related machine learning methods.

\begin{figure}
	\gridline{\fig{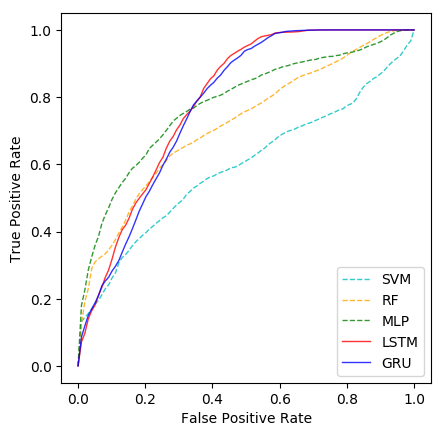}{0.3\textwidth}{(a) T = 12 hr}
		  		\fig{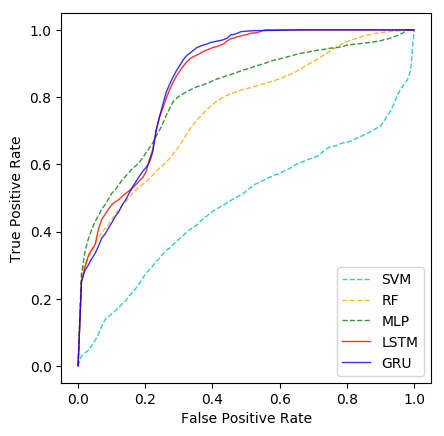}{0.3\textwidth}{(b) T = 24 hr}
				\fig{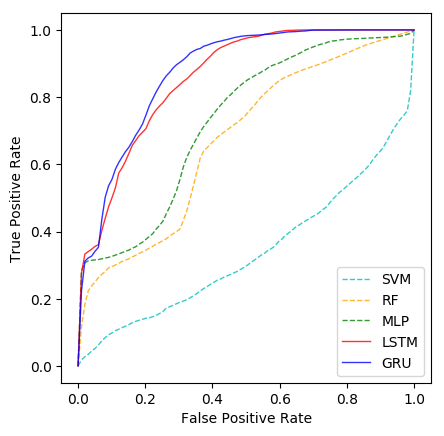}{0.3\textwidth}{(c) T = 36 hr}}
	\gridline{\fig{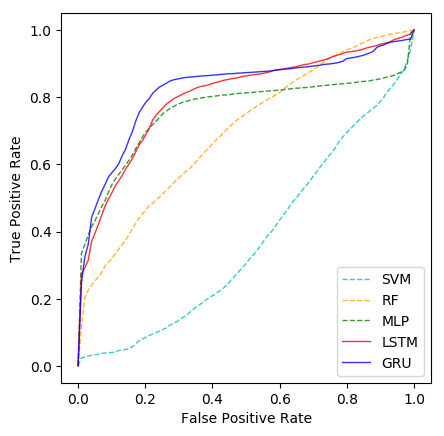}{0.3\textwidth}{(d) T = 48 hr}
		\fig{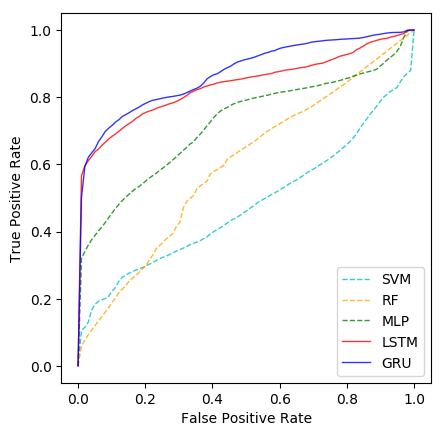}{0.3\textwidth}{(e) T = 60 hr}}
	\caption{ROC curves for our GRU and LSTM networks and three related machine learning methods.}
	\label{fig:roc} 
\end{figure}

Table \ref{tab:confusionmatrix} presents the
confusion matrix in which mean TP, FP, FN, TN 
(with standard deviations enclosed in parentheses)
of the five machine learning methods 
for different $T$ values where $T$ ranges from 12 to 60 hours in 12 hr intervals
are listed.
The probability thresholds used by the machine learning methods
are set to maximize their TSS scores.
Table \ref{tab:comparison} presents the mean performance metric values 
(with standard deviations enclosed in parentheses)
of the five machine learning methods. 
The best performance metric values are highlighted in boldface.
Figure \ref{fig:roc} shows the ROC curves for the five machine learning methods. 
It can be seen from Table \ref{tab:comparison} and Figure \ref{fig:roc} that our GRU and LSTM networks
perform better than the three related machine learning methods
in terms of HSS, TSS, AUC and ROC curves.
There is no clear distinction between the GRU and LSTM networks.
These results indicate that both of the proposed RNNs are suitable 
for solving the binary classification problem at hand.

\begin{table}
	\centering
	\caption{Confusion Matrix for Our GRU and LSTM Networks and Three Related Machine Learning Methods}  
	\label{tab:confusionmatrix}
	\begin{tabular}{cc||c||c||c||c||c}
		\hline
		&              & 12 hr & 24 hr  & 36 hr & 48 hr & 60 hr \\ \hline
		\multirow{5}{*}{TP} & SVM  & 252 (69)  & 138 (153) & 2 (4) & 22 (65) & 112 (141) \\
		& RF & 280 (12) & 580 (13) & 747 (60) & 872 (38) & 883 (78) \\
		& MLP & 395 (15) & 642 (25) & 875 (36) & 1016 (25) & 1188 (87) \\
		& LSTM & 453 (24) & 716 (72) & 964 (57) & 958 (66) & 969 (168) \\
		& GRU & 432 (40) & 770 (32) & 984 (49) & 1107 (60) & 1127 (137) \\ \hline
		
		\multirow{5}{*}{FP} & SVM  & 233 (105) & 110 (142) & 0 (0) & 0 (0) & 72 (144) \\
		& RF & 153 (9) & 314 (10) & 472 (11) & 454 (13) & 485 (9) \\
		& MLP & 270 (33) & 260 (16) & 511 (38) & 295 (27) & 520 (37) \\
		& LSTM & 317 (48) & 292 (41) & 389 (55) & 274 (42) & 49 (50) \\
		& GRU & 294 (73) & 326 (64) & 339 (49) & 262 (42) & 119 (58) \\ \hline
		
		\multirow{5}{*}{FN} & SVM  & 245 (69) & 692 (153) & 1091 (4) & 1364 (65) & 1540 (142) \\
		& RF & 217 (12) & 250 (13) & 345 (60) & 514 (38) & 768 (78) \\
		& MLP & 102 (15) & 188 (25) & 217 (37) & 370 (25) & 464 (88) \\
		& LSTM & 42 (24) & 114 (71) & 128 (57) & 428 (66) & 682 (169) \\
		& GRU & 65 (40) & 60 (32) & 109 (48) & 279 (60) & 524 (138) \\ \hline
		
		\multirow{5}{*}{TN} & SVM  & 455 (105) & 843 (142) & 1127 (1) & 1224 (1) &  1264 (144) \\
		& RF & 535 (9) & 640 (11) & 656 (11) & 770 (13) & 851 (8) \\
		& MLP & 418 (32) & 693 (16) & 617 (38) & 929 (27) & 816 (37) \\
		& LSTM & 369 (48) & 661 (42) & 739 (55) & 951 (42) & 1287 (49) \\ 
		& GRU & 394 (73) & 627 (64) & 789 (49) & 962 (42) & 1216 (57) \\ \hline
	\end{tabular}
\end{table}

\begin{table}
	\centering
	\caption{CME Prediction Results of Our GRU and LSTM Networks and Three Related Machine Learning Methods}  
	\label{tab:comparison}
	\begin{tabular}{cc||c||c||c||c||c}
		\hline
		&              & 12 hr & 24 hr  & 36 hr & 48 hr & 60 hr \\ \hline
		\multirow{5}{*}{Recall} & SVM  & 0.481 (0.142) & 0.197 (0.201) & 0.006 (0.016) & 0.015 (0.044) & 0.390 (0.287) \\
		& RF & 0.627 (0.017) & 0.742 (0.012) & 0.655 (0.054) & 0.680 (0.028) & 0.572 (0.037) \\
		& MLP & 0.783 (0.030) & 0.787 (0.027) & 0.772 (0.036) & 0.727 (0.018) & \textbf{0.733} (0.049) \\
		& LSTM & \textbf{0.945} (0.032) & 0.901 (0.071) & \textbf{0.895} (0.053) & 0.767 (0.046) & 0.644 (0.086) \\
		& GRU & 0.907 (0.064) & \textbf{0.913} (0.048) & 0.883 (0.051) & \textbf{0.815} (0.039) & 0.722 (0.055) \\ \hline
		
		\multirow{5}{*}{Precision} & SVM  & 0.546 (0.086) & 0.380 (0.326) & 0.164 (0.334) & 0.100 (0.300) & 0.705 (0.177) \\
		& RF & \textbf{0.621} (0.016) & 0.641 (0.010) & 0.619 (0.023) & 0.648 (0.013) & 0.642 (0.017) \\
		& MLP & 0.616 (0.024) & 0.708 (0.012) & 0.640 (0.011) & 0.779 (0.013) & 0.694 (0.009) \\
		& LSTM & 0.586 (0.032) & 0.699 (0.020) & 0.710 (0.023) & 0.785 (0.022) & \textbf{0.916} (0.056) \\
		& GRU & 0.598 (0.065) & \textbf{0.712} (0.033) & \textbf{0.754} (0.021)  & \textbf{0.803} (0.020) & 0.884 (0.041)\\ \hline
		
		\multirow{5}{*}{ACC} & SVM  & 0.607 (0.068) & 0.550 (0.039) & 0.511 (0.008) & 0.477 (0.023) & 0.510 (0.068) \\
		& RF & 0.683 (0.011) & 0.686 (0.010) & 0.633 (0.026) & 0.634 (0.016) & 0.587 (0.021) \\
		& MLP & \textbf{0.703} (0.018) & 0.750 (0.011) & 0.673 (0.011) & 0.745 (0.007) & 0.673 (0.018) \\
		& LSTM & 0.694 (0.038) & 0.773 (0.019) & 0.768 (0.024) & 0.765 (0.019) & 0.767 (0.031) \\
		& GRU & 0.696 (0.044) & \textbf{0.786} (0.026) & \textbf{0.800} (0.016)  & \textbf{0.795} (0.016) & \textbf{0.791} (0.015) \\ \hline
		
		\multirow{5}{*}{HSS} & SVM  & 0.182 (0.126) & 0.057 (0.088) & 0.006 (0.016) & 0.014 (0.042) & 0.042 (0.161) \\
		& RF &0.349 (0.022) & 0.376 (0.020) & 0.266 (0.054) & 0.263 (0.032) & 0.176 (0.037) \\
		& MLP &0.413 (0.030) & 0.501 (0.023) & 0.349 (0.023) & 0.491 (0.014) & 0.335 (0.033) \\
		& LSTM & \textbf{0.423} (0.065) & 0.551 (0.039) & 0.537 (0.047) & 0.528 (0.038) & 0.544 (0.056) \\ 
		& GRU & 0.422 (0.075) & \textbf{0.577} (0.050) & \textbf{0.601} (0.033)  & \textbf{0.588} (0.032) & \textbf{0.587} (0.028) \\ \hline
		
		\multirow{5}{*}{TSS} & SVM  & 0.178 (0.125) & 0.056 (0.084) & 0.006 (0.016) & 0.015 (0.044) & 0.049 (0.160) \\
		& RF & 0.350 (0.022) & 0.380 (0.020) & 0.267 (0.054) & 0.262 (0.032) & 0.178 (0.038) \\
		& MLP & 0.428 (0.027) &0.505 (0.023) & 0.350 (0.023) & 0.493 (0.015) & 0.332 (0.031) \\
		& LSTM & \textbf{0.459} (0.066) & 0.562 (0.042) & 0.540 (0.048) & 0.529 (0.037) & 0.562 (0.054) \\
		& GRU & 0.452 (0.068) & \textbf{0.588} (0.049) & \textbf{0.602} (0.033)  & \textbf{0.588} (0.032) & \textbf{0.600} (0.027) \\ \hline
		
		\multirow{5}{*}{AUC} & SVM & 0.592 (0.043) & 0.482 (0.010) & 0.337 (0.072) & 0.373 (0.014) & 0.478 (0.121) \\
		& RF & 0.720 (0.007) & 0.759 (0.006) & 0.670 (0.012) & 0.695 (0.009) & 0.594 (0.009) \\
		& MLP & 0.760 (0.014) & 0.810 (0.016) & 0.728 (0.011) & 0.759 (0.007) & 0.720 (0.033) \\
		& LSTM & \textbf{0.791} (0.023) & \textbf{0.851} (0.014) & 0.858 (0.018) & 0.801 (0.023) & 0.836 (0.019)\\
		& GRU & 0.779 (0.041) & 0.850 (0.017) & \textbf{0.874} (0.016) & \textbf{0.822} (0.009) & \textbf{0.870} (0.021) \\ \hline
	\end{tabular}
\end{table}

\subsection{Probabilistic Forecasting}

The three related machine learning methods are inherently probabilistic forecasting models.
Our proposed RNNs can be easily converted from a binary classification model
to a probabilistic forecasting model as follows.
Instead of comparing the probability, calculated by the sigmoid function in the output layer of the RNNs, with a threshold,
the RNNs simply output the probability.
For a given time point $t$ and an AR that will produce an M- or X-class flare within the next $T$ hours of $t$, 
this output now represents a probabilistic estimate of how likely 
the flare will initiate a CME.

We use the Brier Score (BS) \citep{brier1950verification} 
and the Brier Skill Score (BSS) \citep{wilks2010sampling}
 to quantitatively assess the performance of a probabilistic forecasting model, where
\begin{equation}\label{eq:BS}
\text{Brier Score (BS)}=\frac{1}{N}\sum_{n=1}^{N}(y_n-\hat{y}_n)^2, 
\end{equation}
\begin{equation}\label{eq:BSS}
\text{Brier Skill Score (BSS)} = 1 - \frac{BS}{\frac{1}{N}\sum_{n=1}^{N}(y_n-\overline{y})^2}.
\end{equation}
Here $N$ is the total number of sequences each having $m$ consecutive data samples in the test set, 
$y_{n}$ and $\hat{y}_{n}$ denote the observed probability 
and the estimated probability of the $n$th sequence respectively
as defined in Equation (\ref{eq:loss}), and
$\overline{y}=\frac{1}{N}\sum_{n=1}^{N}y_{n}$ is the average value of all the observed probabilities.
The values of BS range from 0 to 1 with the perfect score being 0.
The values of BSS range from minus infinity to 1 with the perfect score being 1.

Table \ref{tab:probabilistic} presents the mean BS and BSS scores and standard deviations
of the five machine learning methods 
for different $T$ values where $T$ ranges from 12 to 60 hours in 12 hr intervals.
It can be seen from the table that 
the proposed GRU and LSTM networks are comparable, and 
again outperform the three related machine learning methods in terms of BS and BSS.
These results in Table \ref{tab:comparison} and Table  \ref{tab:probabilistic}
suggest that our RNNs work well when used as
either binary classification models or probabilistic forecasting models.

\section{Discussion and Conclusions}
\label{sec:conclusion}
We develop two RNNs, where one is a GRU network and the other is an LSTM network, for CME prediction.
Given a time point $t$ and an AR that produces an M- or X-class flare within the next $T$ hours of $t$ 
where $T$ ranges from 12 to 60 in 12 hr intervals,
our RNNs, when used as binary classification models, can predict whether the AR will also produce a CME associated with the flare.
In addition, our RNNs, when used as probabilistic forecasting models, can produce a probabilistic estimate of 
how likely the M- or X-class flare will initiate a CME. 

We build a dataset of samples, gathered from the JSOC website, in the period from 2010 May to 2019 May;
each data sample has 18 magnetic parameters provided by SHARP. 
We use the data samples during the years of 2010--2014 for training, 
and the data samples during the years of 2015--2019 for testing. 
The training set and test set are disjoint, and hence our RNNs can make predictions on ARs that they have never seen before. 
With extensive experiments, we evaluate the performance of the RNNs 
and compare them with three closely related machine learning methods
including
MLP \citep{Inceoglu2018},
RF \citep{2017ApJ...843..104L} and
SVM \citep{Bobra2016CME}
using different performance metrics. 
All these machine learning methods including ours can be used as binary classification models or probabilistic forecasting models.
The main results are summarized as follows:
\begin{quote}
	1. Solar data samples in an AR are modeled as time series here. 
Unlike the previous method \citep{Bobra2016CME}, which uses one data sample gathered at the time point $t$ to make prediction, 
our RNNs use the $m$ data samples gathered at $t$
and preceding $m-1$ time points to make prediction
($m$ is set to 20 in the study presented here).
To our knowledge, this is the first time that RNNs, which can  
 capture dependencies in the temporal domain of the data samples,
are used for CME prediction.
	
	2. We evaluate the importance of each of the 18 magnetic parameters, or features, used in this study. 
            Our experimental results show that
           using the most important 5-16 features, depending on different $T$ values, 
          can achieve better performance than using all the 18 features together.
        These results are consistent with the findings in the literature, which indicate that
       using fewer, high quality features is often better than using all, including low quality features
     \citep{Alpaydin-2016, Bobra2016CME, DBLP:books/daglib/0040158}.
Developing effective feature ranking and selection procedures has been an active area of research in 
machine learning and related fields.
In general, to find the optimal feature subset among $n$ features, one has to try all 
$2^{n}-1$ combinations of the $n$ features. 
This exhaustive search algorithm becomes impractical when $n$ is large, 
as in our case where $n = 18$.
Consequently, various heuristics based on statistics, randomization, optimization, sampling, clustering,
machine learning, evolutionary computation, genetic algorithms, branch \& bound algorithms
and principal component analysis, to name a few, have been developed 
\citep{Mitra2002, Gevrey2003, GuyonE03, LiuMY04, Seok2004, olden2004, Somol2004, Yoon2005, 
Chandrashekar2014, Xue2016, fisher2018, Tashi2019, Liu2020}. 
In this work, we use the single feature testing heuristic originated from \citep{Laing2012} to rank and select features.  
In related work, \citet{Bobra2016CME} used an F-score heuristic to rank and select features.
Identifying the best feature selection heuristic with the optimal performance in terms of accuracy and execution time 
remains an open problem. 
We plan to further investigate this problem in the future.
	
	3. Our GRU and LSTM networks are comparable; 
there is no clear distinction between them in terms of the performance metrics studied here.
Both of the networks outperform the related machine learning methods including MLP,  RF and SVM
whether they are used as binary classification models or probabilistic forecasting models. 
These findings are based on the data collection scheme in which
data samples in years 2010-2014 are used for training and those in years 2015-2019 are used for testing.
To further understand the behavior of the machine learning methods, we have performed additional experiments as follows.
In each experiment, data samples collected in one of the ten years during the period of 2010-2019
are used for testing and data samples in all the other nine years together are used for training.
There are ten years in the period and hence there are ten experiments.
The average values of the performance metrics are calculated.
The results based on these additional experiments are consistent---our GRU and LSTM networks are comparable, 
and both of them perform better than
the related machine learning methods MLP, RF and SVM.
\end{quote}

\begin{table}
	\centering
	\caption{Probabilistic Forecasting Results of Our GRU and LSTM Networks and Three Related Machine Learning Methods} 
	\label{tab:probabilistic}
	\begin{tabular}{cc||c||c||c||c||c}
		\hline
		&              & 12 hr & 24 hr  & 36 hr & 48 hr & 60 hr \\ \hline
	
		\multirow{5}{*}{BS} & SVM & 0.308 (0.018) & 0.345 (0.012) & 0.347 (0.007) & 0.376 (0.014) & 0.359 (0.007) \\
		& RF & 0.286 (0.003) & 0.295 (0.003) & 0.342 (0.004) & 0.339 (0.002) & 0.400 (0.007) \\
		& MLP  & 0.213 (0.005) & 0.182 (0.005) & 0.210 (0.006) & 0.202 (0.004) & 0.229 (0.015) \\
		& LSTM & \textbf{0.182} (0.105) & \textbf{0.158} (0.007) & 0.165 (0.005) & 0.183 (0.013) & 0.175 (0.010)\\
		& GRU & 0.192 (0.009) & 0.159 (0.004) & \textbf{0.163} (0.006) & \textbf{0.173} (0.005) & \textbf{0.172} (0.005) \\ \hline
		
		\multirow{5}{*}{BSS} & SVM & -0.264 (0.075) & -0.387 (0.048) & -0.390 (0.029) & -0.512 (0.055) & -0.452 (0.028) \\
		& RF  & -0.175 (0.011) & -0.186 (0.011) & -0.367 (0.018) & -0.361 (0.009) & -0.618 (0.026) \\
		& MLP & 0.125 (0.021) & 0.267 (0.019) & 0.160 (0.023) & 0.188 (0.014) & 0.073 (0.062) \\
		& LSTM & \textbf{0.253} (0.043) & \textbf{0.366} (0.027) & 0.340 (0.018) & 0.265 (0.050) & 0.293 (0.039) \\
		& GRU & 0.213 (0.037) & 0.363 (0.018) & \textbf{0.349} (0.024) & \textbf{0.304} (0.021) & \textbf{0.305} (0.022) \\ \hline
	\end{tabular}
\end{table}

Based on our experimental results, we conclude that the proposed RNNs are valid methods for CME prediction.
It should be pointed out that the CME prediction is performed based on the assumption that an M- or X-class flare already exists.
In practice, how does one know whether an AR will produce an M- or X-class flare within the next $T$ hours of some time point $t$?
This question can be answered by using a flare prediction method 
\citep[e.g.,][]{2017ApJ...843..104L, 2018SoPh..293...28F, 2018SoPh..293...48J, 2018ApJ...858..113N, Liu2019}.
Our software package has two components. 
The first component is our previously developed deep learning program \citep{Liu2019}
where one can use the program to predict whether there is an M- or X-class flare within the next $T$ hours of $t$.
If the answer is yes, then one can use the RNN tools developed here to predict whether the flare produces, or is associated with, a CME.
In future work, we plan to further extend the software package to predict other events (e.g., filament eruptions, SEPs).

We thank the referees for very helpful and thoughtful comments.
We also thank the team of \textit{SDO}/HMI for producing vector magnetic field data products. 
The flare catalogs were prepared by and made available through NOAA NCEI.
The CME event records were provided by DONKI.
The related machine learning methods studied here
were implemented in Python.
This work was supported by NSF grant AGS-1927578.
C.L. and H.W. acknowledge the support of NASA under grants  NNX16AF72G, 80NSSC17K0016,
80NSSC18K0673 and 80NSSC18K1705.

\end{document}